\documentclass[twocolumn]{aastex63}

\received{}
\revised{}
\accepted{}
\shorttitle{Taylor Microscale: MMS Turbulence Campaign}
\shortauthors{Bandyopadhyay et al.}
\begin{document}

\title{Direct Measurement of the Solar-Wind Taylor Microscale using MMS Turbulence Campaign Data}

\author[0000-0002-6962-0959]{Riddhi Bandyopadhyay}
\email{riddhib@princeton.edu}
\altaffiliation{Now at: Department of Astrophysical Sciences, Princeton University, Princeton, NJ 08544, USA}
\affiliation{Department of Physics and Astronomy, University of Delaware, Newark, DE 19716, USA}
%\affiliation{Bartol Research Institute, University of Delaware, Newark, DE 19716, USA}	

\author[0000-0001-7224-6024]{William~H. Matthaeus}
\affiliation{Department of Physics and Astronomy, University of Delaware, Newark, DE 19716, USA}
\affiliation{Bartol Research Institute, University of Delaware, Newark, DE 19716, USA}	

\author[0000-0001-8478-5797]{Alexandros Chasapis}
\affiliation{Laboratory for Atmospheric and Space Physics, University of Colorado Boulder, Boulder, Colorado, USA}

\author[0000-0003-1639-8298]{Christopher T. Russell}
\affiliation{University of California, Los Angeles, California 90095-1567, USA}

\author[0000-0001-9839-1828]{Robert J. Strangeway}
\affiliation{University of California, Los Angeles, California 90095-1567, USA}

\author[0000-0001-7188-8690]{Roy B. Torbert}
\affiliation{University of New Hampshire, Durham, New Hampshire 03824, USA}

\author[0000-0001-8054-825X]{Barbara L. Giles}
\affiliation{NASA Goddard Space Flight Center, Greenbelt, Maryland 20771, USA}

\author[0000-0003-1304-4769]{Daniel J. Gershman}
\affiliation{NASA Goddard Space Flight Center, Greenbelt, Maryland 20771, USA}

\author[0000-0001-9228-6605]{Craig J. Pollock}
\affiliation{Denali Scientific, Fairbanks, Alaska 99709, USA}

\author[0000-0003-0452-8403]{James L. Burch}
\affiliation{Southwest Research Institute, San Antonio, Texas 78238-5166, USA}



\begin{abstract}
Using the novel Magnetospheric Multiscale (MMS) mission data accumulated during the 2019 MMS Solar Wind Turbulence Campaign, we calculate the Taylor microscale $(\lambda_{\mathrm{T}})$ of the turbulent magnetic field in the solar wind. The Taylor microscale represents the onset of dissipative processes in classical turbulence theory. An accurate estimation of Taylor scale from spacecraft data is, however, usually difficult due to low time cadence, the effect of time decorrelation, and other factors. Previous reports were based either entirely on the Taylor frozen-in approximation,
which conflates time dependence, or that were obtained using multiple datasets, 
which introduces 
sample-to-sample variation of plasma parameters, or where inter-spacecraft distance were larger than the present study.  The unique configuration of linear formation with logarithmic spacing of the 4 MMS spacecraft, during the campaign, enables a direct evaluation of the $\lambda_{\mathrm{T}}$ from a single dataset, independent of the Taylor frozen-in approximation. A value of $\lambda_{\mathrm{T}} \approx 7000 \, \mathrm{km}$ is obtained, which is about 3 times larger than the previous estimates.  
\end{abstract}



\section{Introduction: Turbulence Scales}\label{sec:intro}
Turbulence is a multi-scale phenomena. The turbulent solar wind possesses structures and processes with broad range of length scales~\citep{Verscharen2019LRSP}. The different characteristic length scales enter into the dynamics in various ways. For example, the correlation scale represents the sizes of the most energetic eddies~\citep{Smith2001JGR}. The mean-free path between collisions determine the collisionality of the plasma. {Proton kinetic physics dominates near the proton inertial length and gyro-radius~\citep{Leamon1998JGR}; similarly electron  physics becomes important at the electron inertial length and gyro-radius~\citep{Alexandrova2012ApJ}}. These different characteristic scales can provide useful information regarding the propagation of energetic particles, such as cosmic rays in the solar wind~\citep{Jokipii1973ARAA}.

Of these various scales there are several related directly to fundamental turbulence properties, and understanding these in various space and astrophysical venues contributes in the  understanding of physical effects ranging from reconnection to particle heating and scattering.
For an initial orientation, We can
appeal to analogies with hydrodynamics, 
to outline
relationships that exist among these 
scales in 
classical turbulence.
Accordingly, we use
as a reference point the case in which the dissipation is controlled by 
a simple scalar kinematic viscosity $\nu$.  
We may begin with the scale 
at which the bulk of turbulence 
energy resides, or is injected; we call this the energy-containing scale $\lambda_{\mathrm{c}}$.
For a turbulence amplitude $Z$, with units of speed, {one finds} immediately a nonlinear time scale, or the eddy turnover time $\tau_{\mathrm{nl}} = \lambda_{\mathrm{c}}/Z$. 
Smaller scale structures will have faster time scales that depend on their characteristics speeds. 
Using Kolmogorov's famous similarity hypothesis as a guide~\citep{Kolmogorov1941a}, we may estimate the speed of structures at smaller scales $\ell$ to be 
$Z_\ell \sim \epsilon^{1/3} \ell^{1/3} = \tau_{\mathrm{nl}} (\ell /L)^{2/3}$, where we also use the \cite{Karman1938PRSL}
estimate of the decay rate $\epsilon \sim Z^3/\lambda_{\mathrm{c}}$. 
probing at still smaller scales,
very much smaller than $\lambda_{\mathrm{c}}$, 
eventually dissipative processes of viscous origin become important. 
As a first approximation 
one may estimate the time scale for dissipation of a structure (e.g., a vortex) at scale $\ell$
to be $\tau_{\mathrm{diss}} = \ell^2/\nu$, using standard viscous dissipation as a model. 
A reasonable way to estimate the characteristic scale at which dissipation becomes dominant is to ask when the eddy or structure at scale $\ell$ become critically damped. 
This occurs when the intrinsic nonlinear time balances the local-in-scale dissipative time.
For this, we solve $\tau_{\mathrm{nl}}(\eta) = \tau_{\mathrm{diss}}(\eta)$ finding 
$\eta = \lambda_{\mathrm{c}} (\nu/Z \lambda_{\mathrm{c}})^{3/4} = \lambda_{\mathrm{c}} Re^{-3/4}$.
The scale $\eta $ is often called the Kolmogorov dissipation scale, and we recognize standard definition of the large scale Reynolds number $Re \equiv { Z \lambda_{\mathrm{c}} / \nu}$.
Note that if the critically damped 
scale is known or estimated, as it might be in a plasma identified with ion inertial scale for example, then the {\it effective Reynolds number} may be defined 
as $R_{\mathrm{eff}} = (\lambda_{\mathrm{c}}/\eta)^{4/3}$, or 
$R_{\mathrm{eff}} = (\lambda_{\mathrm{c}}/d_{\mathrm{i}})^{4/3}$ if dissipation is presumed to become dominant over nonlinear effects at scales comparable to the ion inertial length $d_i$.

Yet another scale, generally intermediate to $\lambda_{\mathrm{c}}$ and $\eta$ may be defined by equating the large scale eddy turnover time to the scale-dependent
dissipative time. Thus,
$\tau_{\mathrm{nl}} = \tau_{\mathrm{diss}}(\ell)$ 
is solved by a particular value $\ell = \lambda_{\mathrm{T}}$. 
This length scale is the \textit{Taylor microscale}, the subject of 
the present paper. 
The first of its several 
equivalent definitions highlights a particular physical property, namely 
that it is critically damped at the large scale nonlinear time. 
Before turning to its evaluation
in the MMS Turbulence Campaign, we 
introduce and discuss several additional properties of the Taylor scale 
$\lambda_{\mathrm{T}}$.  

\section{Taylor Microscale}\label{sec:taylor}
Like the majority of concepts in plasma turbulence, the Taylor scale is also  borrowed from hydrodynamic turbulence research. The Taylor scale can be viewed as the measure of  curvature of the autocorrelation function $\left(R(\mathbf{r}) = \langle \mathbf{F}(\mathbf{x}) \cdot \mathbf{F}(\mathbf{x}+\mathbf{r}) \rangle \right)$ at the origin; for isotropy, 
\begin{eqnarray}
\lambda_{\mathrm{T}}^2 = \frac{R(0)}{R^{\prime \prime}(0)} \label{eq:ltcurv},
\end{eqnarray}
where $\mathbf{F}$ is the fluctuating field of interest, e.g., velocity field $(\mathbf{v})$ in hydrodynamic turbulence, or magnetic field $(\mathbf{b})$ in magnetohydrodynamic (MHD) and plasma turbulence. Here, we consider only the magnetic field fluctuations. For small lags $r$, using $R(\mathbf{r}) = R(-\mathbf{r})$ , a requirement of statistical homogeneity, the autocorrelation
function near the origin can be Taylor expanded, 
assuming isotropy, as
\begin{eqnarray}
R(r) = 1 - \frac{r^2}{2\,\lambda_{\mathrm{T}}^2} + \cdots \label{eq:rapprox1},
\end{eqnarray}

Another physically revealing way to view the Taylor microscale is obtained by noting that for viscous-like $\nu$ dissipation in an incompressible medium, the Taylor scale is also related to dissipation, in that
\begin{eqnarray}
\frac{\mathrm{d} \langle |\mathbf{b}|^2 \rangle}{\mathrm{d} t} = 
\nu \langle |\nabla \times {\bf b}|^2 \rangle =  
\nu \frac{\langle |\mathbf{b}|^2 \rangle}{\lambda_{\mathrm{T}}^2} \label{eq:rdiss}.
\end{eqnarray}

In this sense, the Taylor scale is the ``equivalent dissipation scale," so that, at any instant of time, the dissipation rate is the same if all the energy were at the Taylor
scale. This is, in fact, the physical basis upon which Taylor ~\citep{Taylor1935} first formulated the idea of this particular length scale. Some older turbulence texts~\citep{Hinze} refer to the Taylor scale  as ``the dissipation scale," although later Kolomogrov~\citep{Kolmogorov1941a} introduced 
the similarity variable $(\eta)$, now known as the Kolmogorov length scale,
to 
denote the scale at which 
eddies become critically 
damped.  
Notionally, 
the Taylor microscale represents the largest eddies in the dissipation range, or equivalently, the smallest eddies in the inertial range. The interpretation, of course, may not be so straightforward in plasma turbulence. However, one may draw some conclusions by  analogy with hydrodynamics.

Keeping this parallel in mind, we recall that, indeed, space plasma observations show that the transition of Kolmogorov -5/3 spectra to a steeper slope occur at somewhat larger scales than ion-inertial scale or proton gyro radius~\citep{LeamonEA00}. In classical hydrodynamic turbulence, the Taylor scale is greater than the Kolmogorov length scale $(\eta)$. Therefore, if one treats the ion-inertial length or the proton gyro radius as equivalent to the classical-turbulence Kolmogorov scale, where dissipative (or kinetic) processes become dominant, the Taylor scale provides a natural descriptor of the slight 
steepening of spectra,
and the onset of dissipation, prior to dissipation 
becoming dominant at still smaller scales. This also fits well with the idea, from reconnection studies, that intense kinetic activity in current sheets 
is initiated 
at some multiple of the ion scales \citep{ShayEA98}

\section{Limitations}\label{sec:lim}
The computation of Taylor microscale is, however, challenging in spacecraft data due to low temporal cadence, and other factors, as discussed in the following.
The primary hindrance in accurate estimation of Taylor microscale, using spacecraft data, comes from  the approximation due to Taylor's frozen-in hypothesis~\citep{Taylor1938PRSLA}. In-situ spacecraft data are usually collected in the form of time signals at the spacecraft location. However, usually in the solar wind, the Alfv\'en speed is much smaller than the flow speed, so that one may assume, to a good approximation, that the plasma is frozen-in to the flow \citep{Jokipii1973ARAA}. Therefore, as the plasma convects past the spacecraft, the collected time series data can be essentially interpreted as an one-dimensional ``cut" through the three-dimensional plasma. 
This is the interplanetary spacecraft version of the Taylor's frozen-in wind-tunnel approximation. 
In spite of the widespread utility of this approach, it is apparent that,
when available, 
the correct way to probe spatial structures is through simultaneous multi-point observations~\citep{Matthaeus2019WP}. Moreover, the Taylor approximation is only well-justified for inertial-scale and somewhat larger fluctuations, as 
high frequency kinetic 
activity at sub-proton scales might introduce substantial inaccuracy~\citep{Roberts2020EGU}. 

However, the evaluation of $\lambda_{\mathrm{T}}$, requires 
measurement of 
the curvature of the correlation function \textit{near the origin}, precisely demanding 
such information regarding small spatial range fluctuations. But simultaneous multi-point data, especially at small separation, have generally not been available. Even when multi-point measurements have been obtained, those intervals
are typically not very long~\citep[e.g.,][]{Roberts2015ApJ, Chasapis2017ApJ}. For statistical studies of
turbulence in the solar wind, continuous
intervals of duration of at least a few hours,  corresponding to 
at least a few spacecraft-frame 
correlation times, are desirable~\citep{Isaacs2015JGR}. Consequently, previous studies were forced to perform analyses 
compiled from a number of different intervals. At that point, additional uncertainty is introduced by 
variation of plasma conditions 
from interval-to-interval. 
The first MMS mini campaign, named the \textit{MMS Solar-Wind Turbulence Campaign}, explicitly overcomes these limitations, as discussed in the next section. {However, we note that the data studied are also limited in that there is one sampling direction i.e., all spacecraft are in a line. Although four-point measurements provide many advantages, this and other related limitations are intrinsic to four-point measurements.}

%\clearpage
\section{MMS Solar-Wind Turbulence Campaign}\label{sec:camp}
\begin{figure}
	\begin{center}
		\includegraphics[width=\linewidth]{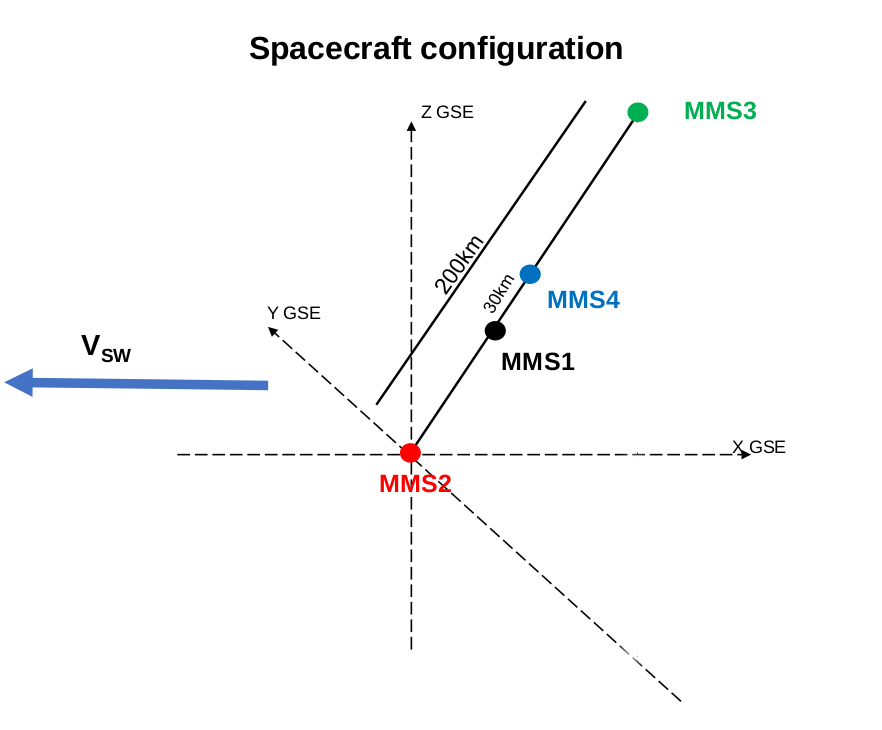}
		\includegraphics[width=0.8\linewidth]{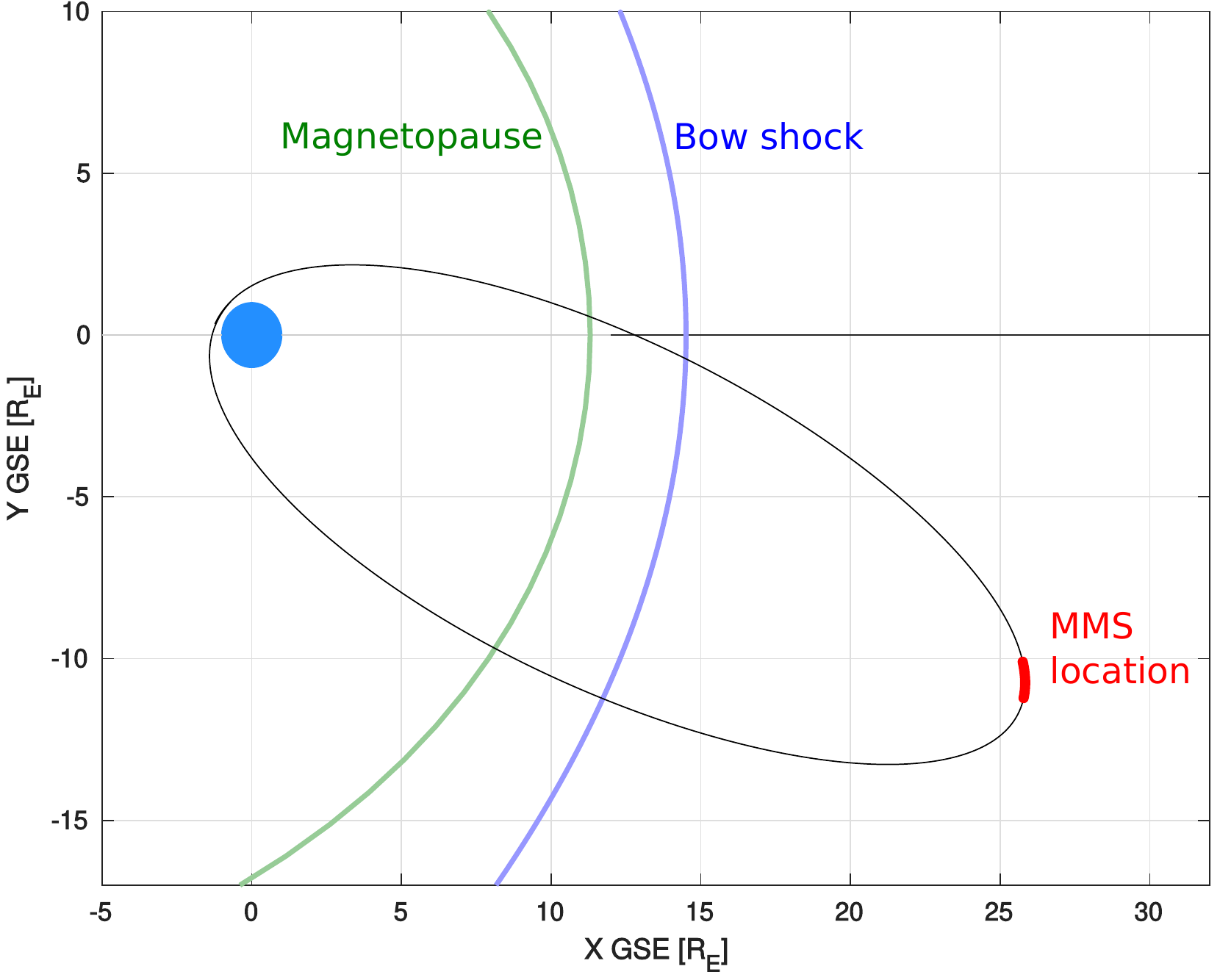}	
		\caption{Top: Configuration of the 4 MMS  spacecraft and the flow speed direction during the MMS turbulence campaign. Bottom: Location of the MMS, along with the nominal magnetopause and bow-shock locations. The
Geocentric Solar Ecliptic (GSE)~\citep{Franz2002PSS} coordinate system is used, in which the XY-plane is defined by the Earth mean
ecliptic and the +X-axis is defined by the Earth-Sun vector.}
		\label{fig:camp}
	\end{center}
\end{figure}

The Magnetospheric Multiscale (MMS) mission, was launched in 2015 with the primary goal of  studying magnetic reconnection — a process responsible for releasing 
magnetic energy into flows and internal energy. The four MMS spacecraft are equipped with state-of-the-art instruments with unprecedented resolution. During February 2019, the MMS apogee was raised to $\sim 27\,\mathrm{R_{E}}$ on Earth's
dayside of the magnetosphere and outside the ion foreshock
region. This orbit allowed the spacecraft to sample the pristine
solar wind, outside the Earth’s magnetosheath and far from the bow shock, for extended
periods of time (see Fig.~\ref{fig:camp}).

During the first mini campaign, the four MMS spacecraft were arranged in a ``string of pearls" or ``beads on string" formation instead of the usual tetrahedral formation. 
With the spacecraft baseline almost perpendicular to the solar wind flow, the spacecraft were separated by logarithmic distances ranging from 25 to 200 km, {and the baseline separations remain unchanged within $10\%$.} This configuration allows direct investigation
 of the scale-dependent nature of the solar wind structures near proton scales. Laboratory experiments have utilized such formations~\citep{Cartagena-Sanchez2019AIP}, but this kind of data are novel in observations. This work is the first of several studies undertaken to take advantage of this unique configuration (see also \citet{ChasapisEA2020}).
Although not relevant for this study, the spacecraft spin-axis were tilted about 15$^\circ$ to obtain improved electric field measurements in the solar wind. A schematic configuration 
of the four MMS spacecraft
in the solar wind, during the Turbulence Campaign, 
is provided in the top panel of Fig.~\ref{fig:camp}. 
The orbital context plot showing 
MMS location relative to the nominal magnetopause~\citep{Shue1998JGR} and bow shock~\citep{Farris1994JGR}, is illustrated in the bottom panel of Fig.~\ref{fig:camp}.

%\clearpage

\section{MMS Data}\label{sec:data}
\begin{figure}
	\begin{center}
		\includegraphics[width=\linewidth]{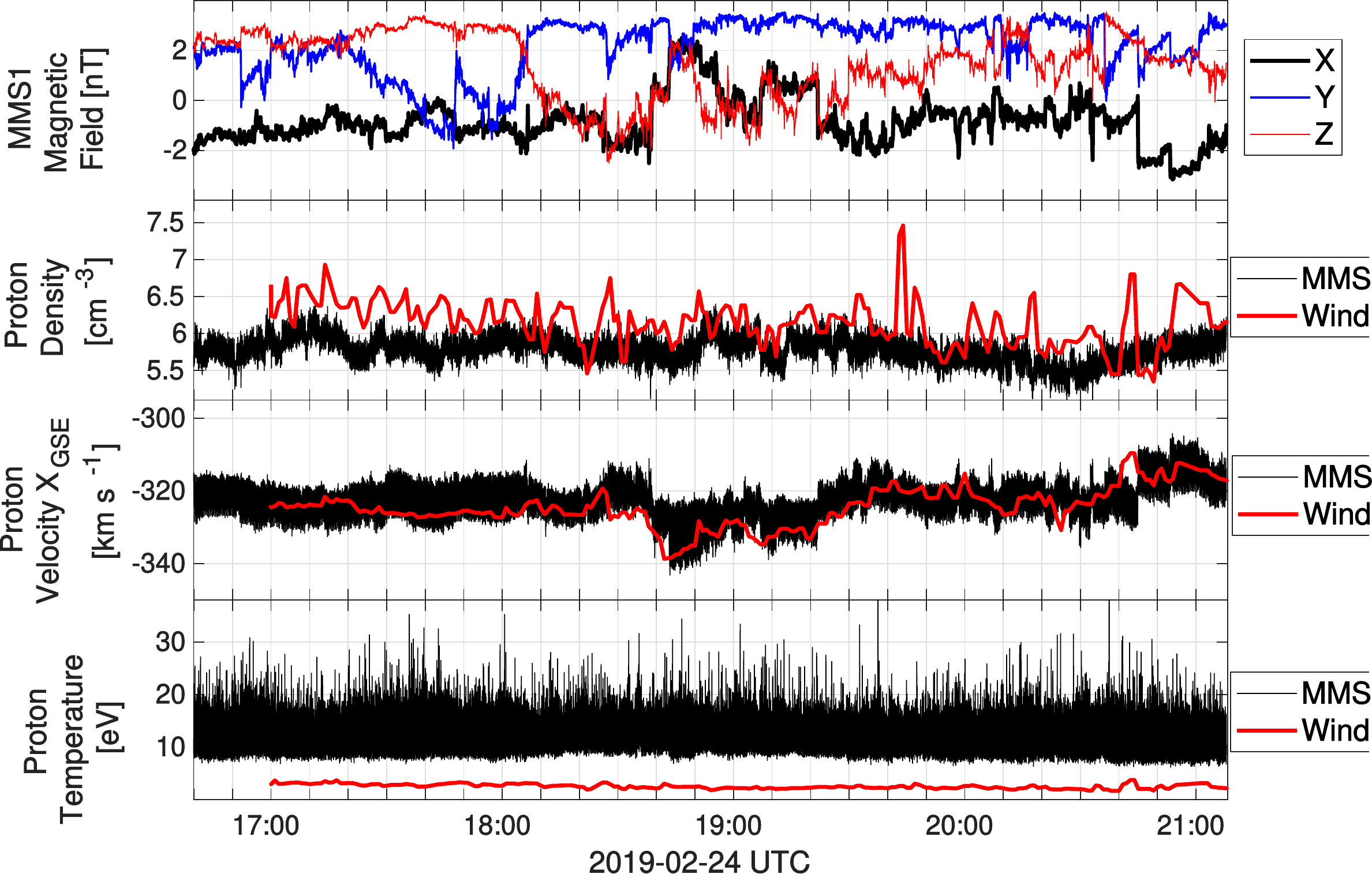}
		\caption{Time series of solar wind observations from 16:39:00 - 21:41:00 UTC on 24 Feb 2019. Top panel: GSE components ofthe MMS1/FGM magnetic field, second panel: Proton number density measured by MMS1 and Wind, third panel: X component of the proton velocity, in GSE coordinate system, measured by MMS1 and Wind, bottom panel: Proton temperature measured by MMS1 and Wind.}
		\label{fig:data}
	\end{center}
\end{figure}

\begin{table}
\begin{center}
\caption{Parameters for MMS interval on 
24 February 2019, from 16:00 to 21:00 UTC (5 hours). Quantities with an asterisks $(^*)$ have been estimated using Wind data, and for those, the MMS estimates are given in parenthesis.}
\label{tab:parameters}
\begin{tabular}{l|c}
Solar-wind speed  & $V_{\mathrm{SW}}=$ 322 km s$^\mathrm{-1}$\\
Correlation Length & $\lambda_{\mathrm{c}}=3.2 \times 10^5$ km \\
Ion inertial length & $d_{\mathrm{i}}=$ 91 km\\
Ion gyroradius & $\rho_{\mathrm{i}}=$ 64$^*$ (150) km \\
Electron inertial length & $d_{\mathrm{e}}=$ 2.3 km\\
Debye length & $\lambda_{D} =$ 10 m \\
Proton beta & $\beta_{\mathrm{i}}=$ 0.5$^{*}$  (2.5) \\
Magnetic field & $B_0 = |\langle \mathbf{B} \rangle|=$ 3.4 nT\\
Magnetic-field fluctuation & $B_{\mathrm{rms}}/B_0=$ 0.72\\
Proton density & $\langle N_{\mathrm{i}} \rangle= $ 6.2 cm$^{-3}$\\
Proton temperature  & $\langle T_{\mathrm{i}} \rangle=$ 2.5$^*$  (12.4) eV\\
\end{tabular}
\end{center}
\end{table}

During the three-week long mini campaign, a number of useful solar wind and foreshock intervals were selected. The longest of the selected pristine solar wind intervals, a continuous interval of five hours of burst-mode data, on 24 February 2019, from 16:39:00 to 21:41:00 UTC, 
is analyzed in this paper. No signature of reflected ions from the bow shock is found. {For this interval, we did not detect any high-frequency waves characteristic of the foreshock. We note, however, that the other 5-hour interval (17 February 2019, from 11:24:00 to 16:24:00 UTC) chosen as a part of the turbulence campaign, has foreshock signatures, and consequently that interval was not considered for this analysis.}

To evaluate the magnetic field  
Taylor microscale from two-spacecraft correlation data, 
We employ data from the Fluxgate Magnetometer (FGM) aboard each 
each of the four MMS spacecraft\citep{Russell2016SSR}. The top panel of  Fig.~\ref{fig:data} shows the three Cartesian components 
of magnetic field in the GSE coordinate system~\citep{Franz2002PSS},
recorded in this period by the FGM onboard MMS1. It is apparent that this interval is rich in structures, including numerous current sheets, flux tubes, and broad band random fluctuations - taken together 
these represent a fairly typical 
sample of solar wind turbulence~\citep{Bruno2005LRSP}.

The Fast Plasma Investigation (FPI)~\citep{Pollock2016SSR} instrument measures proton and electron distribution functions  and moments every 150ms and 30ms, respectively.
Due to the limitations of the FPI instruments in the solar wind~\citep{Bandyopadhyay2018aApJ}, some systematic
uncertainties remain in the moments,
and more so in the higher-order moments.
{Therefore, we cross-check the proton moments in the
selected interval with Wind~\citep{Ogilvie1995SSR, Lepping1995SSR} data, time-shifted to the bow-shock nose.} The MMS and Wind estimates of proton density, velocity (X$_{\mathrm{GSE}}$ component), and temperature are shown in the bottom three panels in Fig.~\ref{fig:data}.
The density and velocity are in adequately close agreement, but significant discrepancies exist in the proton temperature values. The FPI estimates of temperature are significantly greater than the Wind values. Given the known limitations of FPI in the solar wind, we use the Wind measurements of temperature to evaluate  proton beta and other relevant parameters. The average values of the plasma parameters are reported in Table~\ref{tab:parameters}.

%\clearpage

\begin{figure}
	\begin{center}		
	\includegraphics[width=\linewidth]{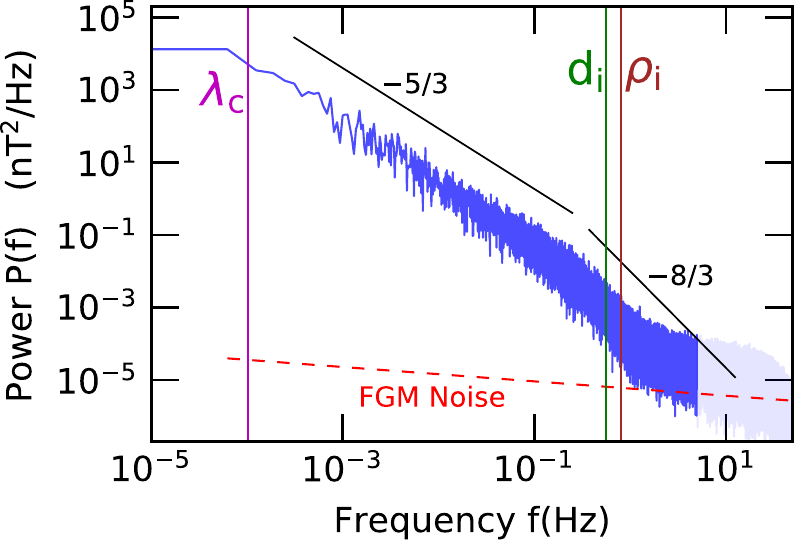}
	\caption{Spectral power density of magnetic field measured by MMS1.
Kolmogorov scaling $\sim f^{-5/3}$ is shown for reference. The vertical lines
represent the correlation length $(k \lambda_{\mathrm{c}}=1)$, the ion-inertial length $(k d_{\mathrm{i}}=1)$, and the ion gyro-radius $(k \rho_{\mathrm{i}}=1)$, with wavenumber $k \simeq (2 \pi f)/|\langle \mathbf{V} \rangle|$. The part of the spectrum where the signal-to-noise ratio decreases below $\sim 5$, is grey-shaded, to indicate that this region is noise dominated. {Note that the flattening in the high-frequency range ($f \gtrsim 1$ Hz) is due to noise and not physical (see text).}}
	\label{fig:spec}
	\end{center}
\end{figure}

Fig.~\ref{fig:spec} shows spacecraft-frame 
frequency spectrum of the magnetic field during this period. A clear Kolmogorov scaling $(\sim f^{-5/3})$ can be seen at scales smaller than the correlation length, $\lambda_{\mathrm{c}}$
(inferred from the Taylor hypothesis). A break in
spectral slope from $\sim f^{-5/3}$ to $\sim f^{-8/3}$ is observed near (inferred) kinetic scales ($d_{\mathrm{i}}$ or $\rho_{\mathrm{i}}$). Often these scales are associated with the dissipation scale $(\lambda_{\mathrm{diss}})$ in collisionless plasmas, equivalent to Kolmogorov scale $(\eta)$ in classical turbulence. Kinetic dissipative processes, such as wave damping, are effective in these small plasma kinetic scales. For example, \cite{LeamonEA00} 
and \cite{Wang2018JGR}
argued that the ion inertial scale 
controls the spectral break and onset of strong  dissipation, while 
 \cite{Bruno2014ApJL} suggested the
 break frequency is associated with the resonance condition for parallel propagating Alfv\'en wave.
 Another possibility is that the largest of the proton kinetic scales terminates the inertial range and controls 
 the spectral break\citep{ChenEA14-grl}. {The flattening near $f \gtrsim 1$ Hz is very likely to be noise dominated, since, for example, this behavior is not seen in Cluster search coil observations~\citep[e.g.,][]{Alexandrova2009PRL, Alexandrova2012ApJ, Roberts2017ApJ}.}
 
\section{Taylor Microscale: Results}\label{sec:results}

\begin{figure}
	\begin{center}
	\includegraphics[width=\linewidth]{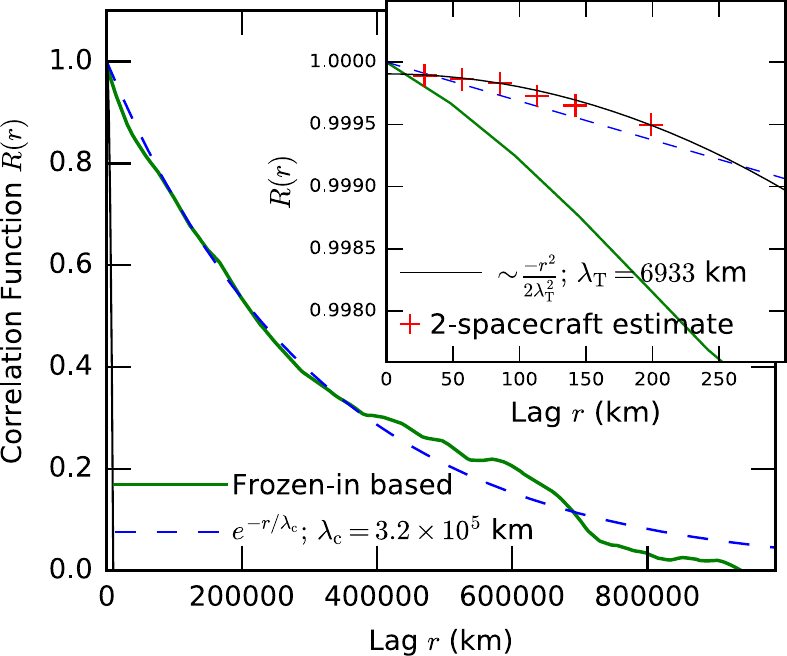}
		\caption{Magnetic-field correlation function based on frozen-in approximation (green, solid line) and obtained from two-spacecraft evaluation (red, cross symbols). An exponential fit (blue, dashed line) to the single-spacecraft measurement is used to obtain the correlation length. A quadratic fit (black, thin line) to the multi-spacecraft points estimates the Taylor scale. The inset plots part of the correlation function enlarged near the origin to clearly show the multi-spacecraft points and the parabolic fit.}
		\label{fig:taylor}
	\end{center}
\end{figure}

To estimate the Taylor scale from this interval, we recall the approximation near the origin:
\begin{eqnarray}
R(r) \approx 1 - \frac{r^2}{2\,\lambda_{\mathrm{T}}^2}  \label{eq:rexpa2},
\end{eqnarray}
where the higher-order terms are neglected. Therefore, one may obtain the Taylor microscale by fitting the autocorrelation function $R(r)$ to a parabolic curve at the origin. Clearly, the quadratic approximation holds better as one asymptotically 
approaches smaller values of $r$. Previous multi-spacecraft estimates~\citep{Matthaeus2005PRL} were evaluated with the Cluster spacecraft, with separations in the range 150 km $ \leq r \leq $ 270 km. Here, we extend that analysis
by approaching the 
origin closer by 
about an order of magnitude, 
with 25 km $ \leq r \leq $ 200 km. {Appendix~\ref{sec:error} provides an estimate of the accuracy of the correlation measurements.} As shown in Fig.~\ref{fig:taylor}, we extract an
estimate of $\lambda_{\mathrm{T}}$ by fitting the 
the six available two-point correlation function to a parabolic curve. The resulting value of the Taylor scale is $\lambda_{\mathrm{T}}=6933$km. For comparison, we also show the single-spacecraft, frozen-in hypothesis based evaluation of the correlation function. Evidently, the single-spacecraft estimate decays much rapidly closer to the origin, presumably due to time decorrelation of the solar wind fluctuations in those scales~\citep{MatthaeusEA10}. 
At large lags, however, the frozen-in based correlation function exhibits approximately exponential decay and provides a satisfactory estimate of the correlation length, about $320,000$km, 
consistent with previous reports~\citep{Isaacs2015JGR}.
Note that the exponential
passes close to the multi-spacecraft points (see inset) near  the origin; however, the exponential cannot be employed to determined $\lambda_{\mathrm{T}}$ since the curvature at the origin is undefined. 

\textit{Solar wind viscosity}-- 
An accurate estimation  of the Taylor scale also permits an 
evaluation of an effective viscosity 
(or, 
turbulent viscosity/resisivity) of the solar wind, according to the expression, 
\begin{equation}
\nu = \frac{\epsilon}{2E} \lambda_T^2, \label{eq:visc}
\end{equation}
where $\epsilon$ is the cascade rate $\approx 1000\,\mathrm{J\,kg\,s^{-1}}$  
and 
$E$ is the fluctuation energy per unit mass. 
 The cascade rate can be obtained, for example, 
 from the third order law or other estimates, see e.g., \cite{Verma1995JGR, Sorriso-Valvo2007PRL, MacBride2008ApJ, Bandyopadhyay2018bApJ}. Putting in the rest of the values: $\lambda_{\mathrm{T}} = 6933$ km and $2 E = \langle |\mathbf{b}|^2 \rangle = 324\,\mathrm{km}^2\,\mathrm{s}^{-2}$, we obtain $\nu \approx 150\,\mathrm{km}^2\,\mathrm{s}^{-1}$. This value is considerably larger than the one obtained using Braginskii~\citep{Braginskii1965RPP} formalism, which is based on simple particle-particle collision~\citep{Montgomery1983SW5}. Our result also provides improvement on earlier indirect estimates, based on turbulence-cascade phenomenology~\citep{Coleman1968ApJ, Verma1996JGR}.

\section{Discussion and Conclusions}\label{sec:disc}
In this paper, 
MMS data accumulated during the turbulence campaign have been used to evaluate the Taylor microscale of magnetic field fluctuations using a multi-spacecraft technique,
and taking advantage of a unique beads-on-a-string flight formation. The previous estimate by \cite{Matthaeus2005PRL}, using Cluster data, is $\lambda_{\mathrm{T}}=2478$ km, which is about 3 times smaller than the present evaluation. The deviation is possibly due to the relatively larger spacecraft separation used in the Cluster data set, comparatively shorter intervals, and mixing of different solar wind intervals. 
It is also possible that this level of variability is intrinsic to the solar wind for a variety of reasons including $1/f$ noise, stream structure~\citep{Matthaeus1986PRL}. {Another possibility is that the differences may be attributable to the differences in the formation of the Cluster and MMS spacecraft.  The Cluster spacecraft pair baselines are in a tetrahedron, introducing anisotropy effects, which are not present in the MMS linear formation analyzed here. These limitations are inherent to  four-point measurements, and can be overcome by large constellations of simultaneous, in-situ measurements~\citep{Matthaeus2019WP, Klein2019WP, TenBarge2019WP}.}

{We note here that the two-spacecraft data points cover a very small range, in contrast with the frozen-in based correlation function (compare the inset in Fig.~\ref{fig:taylor} to the main plot). We, however, do not expect any weakness in the analysis due to this point. The estimation of correlation length by an exponential approximation is only valid at long lag, while the quadratic approximation to the expansion of the  correlation function  is expected to hold only near the origin. Therefore, the small coverage of scale is not expected to hinder the Taylor-scale estimation.}

We find that the break frequency, in the magnetic-field spectrum, is situated between the Taylor scale $(\lambda_{\mathrm{T}})$ and dissipative scales $(d_{\mathrm{i}}, \rho_{\mathrm{i}})$. {\cite{Wang2018JGR}, showed that for $\beta \sim 1$ plasma, the spectral break frequency is better associated with $d_{\mathrm{i}}$ than $\rho_{\mathrm{i}}$, and it is insensitive to 
$\beta$-values.} In general, $\lambda_{\mathrm{T}} / \lambda_{\mathrm{diss}} > 1$ in hydrodynamic turbulence,
and the separation increases with Reynolds number. Although the dissipation mechanism in the collsiionless solar wind is not due to a viscous closure, we note here that if one associates the ion-inertial length $d_\mathrm{i}$ or ion gyro-radius $\rho_\mathrm{i}$ with the dissipation scale, then we find that $\lambda_{\mathrm{T}} / \lambda_{\mathrm{diss}} \approx 70$.
The relationship of Taylor scale to ion kinetic
scales in the solar wind however, appears to be much more variable than it is in hydrodynamics, and in particular the relationship has been found to depend on the turbulence cascade rate 
\citep{MattEA08-Taylor}.

The present paper is a step in a broad progression of interplanetary measurements of fundamental 
plasma turbulence properties. The 1980 NASA Plasma Turbulence Explorer Panel emphasized the need for simultaneous multi-point measurements, in particular, plasma and magnetic field measurements, to make progress in this area~\citep{Montgomery1980RNPTES}. More recently, the space plasma community has witnessed growing interest in understanding the multi-scale nature of turbulence processes in space, using multi-point measurements in observations~\citep{Matthaeus2019WP,Klein2019WP} as well as laboratory experiments~\citep{Schaffner2014ApJ, Schaffner2015ApJ}, especially with regard to the physics of dissipation and heating mechanisms. The MMS turbulence campaign provides the first opportunity to capture multi-scale processes near the proton scales
with a single data interval. Therefore, the results presented in this paper will be useful for future and proposed multi-spacecraft  missions~\citep{Vaivads2016JPP, Bookbinder2018AGU, Plice2019helioswarm, Verscharen2019EGU, Wicks2019EGU}.

The 2019 Solar Wind turbulence campaign is the first of the many MMS mini-campaigns that are planned to be held in the second extended  mission phase. The results presented in this paper will serve as a demonstration of the MMS instrumental capabilities, working outside their original region of interest. Exploring the potential advantages of different MMS formations will allow better understanding of MMS range of capabilities, 
which will open the door to other  scientific campaigns.

\section*{Acknowledgments}
This research is partially supported by NASA 
under the MMS mission
Theory and Modeling Team 
grants NNX14AC39G and 80NSSC19K0565, 
by Heliophysics Supporting Research 
grants NNX17AB79G and 80NSSC18K164880,
and by Helio-GI grant NSSC19K0284.
We thank Bob Ergun for initiating 
the MMS Solar Wind Turbulence Campaign and the SITL selection team, including Tai Phan, Benoit Lavraud, Sergio Toledo-Redondo, Julia Stawarz, Rick Wilder, and Olivier Le Contel for helping to select several Solar Wind intervals during the campaign. We are grateful to the MMS instrument teams,
especially SDC, FPI, and FIELDS, for cooperation and
collaboration in preparing the data. The data used in this
analysis are Level 2 FIELDS and FPI data products, in
cooperation with the instrument teams and in accordance their
guidelines. All MMS data are available at \url{https://lasp.colorado.edu/mms/sdc/}. The Wind data, shifted to
Earth's bow-shock nose, can be found at \url{https://omniweb.gsfc.nasa.gov/}. The authors thank the Wind team for the proton moment dataset.

\appendix

\section{Measure of Confidence in the two-spacecraft analyses}\label{sec:error}
{As can be seen in Fig.~\ref{fig:taylor}, the signals between spacecraft are strongly correlated; however, the differences are very small between them. The very small variation in the 2-spacecraft correlations, between $1$ and $0.9995$, calls for further analyses of its accuracy. Here, we provide an error estimate to check the quality of the measurements.

Recall the definition of correlation function for magnetic-field fluctuations $\mathbf{b}$,
\begin{eqnarray}
R(\mathbf{r}) = \langle \mathbf{b}(\mathbf{x}) \cdot \mathbf{b}(\mathbf{x}+\mathbf{r}) \rangle \label{eq:corr}.
\end{eqnarray}
Keeping in mind that the uncertainty in FGM magnetic field measurements are less than $\delta b = 0.1$ nT~\citep{Russell2016SSR}, and that the average magnetic field is about $b \sim 2$ nT, the fractional error in the individual sample points of the correlation function, $\mathbf{b}(\mathbf{x}) \cdot \mathbf{b}(\mathbf{x}+\mathbf{r})$, is less than $2 \delta b / b \sim 0.1$. Further, averaging over all the data points reduces the statistical error by $\sim 1/\sqrt{n}$, where $n$ is the number of data points (The error in the mean estimate would be even smaller by another factor of $1/\sqrt{n}$). So we estimate the error in the 2-spacecraft $R(r)$ as $\delta R / R \sim 10^{-7}$. Therefore, even by a very conservative estimate, the 2-spacecraft values are reliable up to $6$ decimal points. Finally, the root-mean-square of two-spacecraft magnetic-field increments, for the smallest separation, is about 5 times larger than the noise level at the scale of spacecraft separation~\citep[also see ][]{Chhiber2018JGR}, so that we expect that the 2-spacecraft signals are larger than the instrumental noise. These tests indicate that the two-spacecraft analyses are indeed reliable.
}

%\bibliography{refs_riddhi}{}

\bibliographystyle{aasjournal}

\end{document}